\documentclass[pdftex,pagesize=auto,paper=letter,11pt,abstract=on,bibliography=oldstyle]{scrartcl}\usepackage[utf8]{inputenc}\usepackage{amsmath,amssymb}\usepackage[dvipsnames]{xcolor}\usepackage{booktabs,textcomp,rotating}\usepackage{xspace}\usepackage{tikz}\usepackage[bookmarks=false]{hyperref}\usepackage{wrapfig}\usepackage[sort,compress]{cite}\usepackage[format=plain,labelformat={simple},labelsep={period},labelfont={bf}]{caption}\usepackage{authblk}\definecolor{kit-gruen}{cmyk/RGB}{1,0,.6,0/0,150,130} \definecolor{kit-gruen70}{cmyk/RGB}{.7,0,.42,0/76,181,167} \definecolor{kit-gruen50}{cmyk/RGB}{.5,0,.3,0/127,202,192} \definecolor{kit-gruen30}{cmyk/RGB}{.3,0,.18,0/178,223,217} \definecolor{kit-gruen15}{cmyk/RGB}{.15,0,.09,0/217,239,236}\definecolor{kit-blau}{cmyk/RGB}{.8,.5,0,0/70,100,170} \definecolor{kit-blau70}{cmyk/RGB}{.56,.35,0,0/125,146,195} \definecolor{kit-blau50}{cmyk/RGB}{.40,.25,0,0/162,177,212} \definecolor{kit-blau30}{cmyk/RGB}{.24,.15,0,0/199,208,229} \definecolor{kit-blau15}{cmyk/RGB}{.12,.075,0,0/227,232,242}\definecolor{kit-schwarz}{cmyk/RGB}{0,0,0,1/0,0,0} \definecolor{kit-schwarz70}{cmyk/RGB}{0,0,0,.7/77,77,77} \definecolor{kit-schwarz50}{cmyk/RGB}{0,0,0,.5/128,128,128} \definecolor{kit-schwarz30}{cmyk/RGB}{0,0,0,.3/179,179,179} \definecolor{kit-schwarz15}{cmyk/RGB}{0,0,0,.15/217,217,217}\definecolor{kit-maigruen}{cmyk/RGB}{.6,0,1,0/92,172,53} \definecolor{kit-gelb}{cmyk/RGB}{0,.05,1,0/225,227,18} \definecolor{kit-orange}{cmyk/RGB}{0,.45,1,0/229,131,35} \definecolor{kit-braun}{cmyk/RGB}{.35,.5,1,0/144,105,43} \definecolor{kit-rot}{cmyk/RGB}{.25,1,1,0/155,23,35} \definecolor{kit-lila}{cmyk/RGB}{.25,1,0,0/152,5,104} \definecolor{kit-cyan-blau}{cmyk/RGB}{.9,.05,0,0/0,144,204} \usetikzlibrary{automata,positioning,arrows,calc,shapes,decorations.text,decorations.pathmorphing,external,shapes.callouts,decorations.markings}\pgfdeclarelayer{edgelayer} \pgfdeclarelayer{nodelayer} \pgfsetlayers{edgelayer,nodelayer,main} \tikzset{initial text={}} \tikzset{>=stealth'} \tikzset{shorten >=1pt} \tikzset{shorten <=1pt} \tikzstyle{every pin}=[pin distance=.25cm]\tikzstyle{every state}=[minimum size=20pt,draw=kit-schwarz,fill=kit-gruen15] \tikzstyle{loop transition}=[->,kit-schwarz] \tikzstyle{link transition}=[->,kit-blau,thick,every node/.style={fill=white}] \tikzstyle{vertex}=[style={circle,draw=kit-schwarz,fill=kit-gruen15}] \tikzstyle{deleted vertex}=[vertex,style={fill=kit-rot!15}] \tikzstyle{highlighted edge}=[draw,line width=4pt,-,kit-gruen30] \tikzstyle{squiggly edge}=[draw,decorate,decoration={snake,segment length=15pt}] \tikzstyle{squiggly arc}=[draw,->,decorate,decoration={snake,segment length=20pt,post length=3pt}] \tikzstyle{smallvertex}=[style={circle,fill=kit-schwarz70,inner sep=0pt,minimum size=8pt}] \tikzstyle{tinyvertex}=[style={circle,fill=kit-schwarz70,inner sep=0pt,minimum size=5pt}] \tikzstyle{directed edge}=[->]\tikzstyle{plotpoint}=[style={circle,draw,thick,kit-gruen,fill=white,inner sep=0pt,minimum size=5pt}] \tikzstyle{paretopoint}=[plotpoint,style={fill=kit-gruen}]\tikzstyle{arrival vertex}=[style={circle,draw=kit-schwarz,fill=kit-schwarz!0}] \tikzstyle{departure vertex}=[style={circle,draw=kit-schwarz,fill=kit-schwarz}] \tikzstyle{transfer vertex}=[style={circle,draw=kit-schwarz,fill=kit-schwarz!30}] \tikzstyle{landmark vertex}=[vertex,fill=kit-gelb!50] \tikzstyle{stay arc}=[->,>=stealth'] \tikzstyle{departure arc}=[->,>=stealth'] \tikzstyle{connection arc}=[->,>=stealth',thick,every node/.style={fill=white}]\tikzstyle{stop vertex}=[style={circle,draw=kit-schwarz,fill=kit-blau!50}] \tikzstyle{airport vertex}=[style={circle,draw=kit-schwarz,fill=kit-blau!50}] \tikzstyle{superstop vertex}=[style={rectangle,rounded corners,draw=kit-schwarz,fill=kit-blau!50,minimum size=12pt}] \tikzstyle{sourcetargetsuperstop vertex}=[style={rectangle,rounded corners,draw=kit-schwarz,fill=kit-braun!15,minimum size=12pt}] \tikzstyle{hubsuperstop vertex}=[style={rectangle,rounded corners,draw=kit-schwarz,thick,fill=kit-blau!50,minimum size=12pt}] \tikzstyle{viasuperstop vertex}=[style={rectangle,rounded corners,draw=kit-schwarz,thick,fill=kit-rot!50,minimum size=12pt}] \tikzstyle{route vertex}=[style={circle,draw=kit-schwarz,fill=kit-orange!50,minimum size=10pt}] \tikzstyle{transfer arc}=[->,>=stealth',bend right] \tikzstyle{airport arc}=[->,>=stealth'] \tikzstyle{route arc}=[->,>=stealth',thick,every node/.style={fill=white}] \tikzstyle{squiggly route arc}=[->,thick,draw,decorate,decoration={snake,segment length=20pt,post length=5pt}]\tikzstyle{conflict vertex}=[style={rectangle,rounded corners=3pt,draw=kit-schwarz,fill=kit-gruen15,minimum size=10pt}] \tikzstyle{conflict edge}=[-] \tikzstyle{partition boundary}=[-,very thick,kit-schwarz30] \tikzstyle{cut edge}=[thick,-,kit-rot] \tikzstyle{clique edge}=[-,kit-gruen] \tikzstyle{skeleton vertex}=[tinyvertex,kit-gruen]\tikzstyle{stop}=[style={circle,very thick,draw=white,inner sep=0pt,minimum size=6pt}] \tikzstyle{importantstop}=[style={circle,very thick,draw=white,inner sep=0pt,minimum size=9pt}] \tikzstyle{transferstop}=[style={circle,draw=kit-schwarz,inner sep=0pt,minimum size=6pt,fill=white}] \tikzstyle{labeledstop}=[style={rectangle,rounded corners=3pt,draw=kit-schwarz,fill=white}] \tikzstyle{sourcetargetstop}=[style={rectangle,rounded corners=3pt,draw=kit-schwarz,fill=kit-braun!15}] \tikzstyle{route}=[-,line width=3pt,draw,rounded corners=3pt] \tikzstyle{walking}=[line width=1.5pt,kit-schwarz!90] \tikzstyle{cycling}=[line width=1.5pt,kit-lila] \tikzstyle{squigglyroute}=[-,line width=3pt,draw,decorate,decoration={snake,segment length=20pt}]\pgfmathsetseed{3141592} \usepackage[T1]{fontenc}\usepackage{lmodern}\usepackage{microtype}\newcommand{\enabled}{\ensuremath{\bullet}}\newcommand{\disabled}{\ensuremath{\circ}}\newcommand{\feature}[1]{\begin{rotate}{60}\hspace{-.33ex}#1\end{rotate}}\newcommand{\Xcomment}[1]{}\newcommand{\musec}{\textmu{}s\xspace}\newcommand{\wrt}{w.\,r.\,t.\xspace}\newcommand{\eg}{e.\,g.\xspace}\newcommand{\ie}{i.\,e.\xspace}\newcommand{\cf}{cf.\xspace}\newcommand{\depTime}{\ensuremath{\tau}\xspace}\newcommand{\astop}{\ensuremath{p}\xspace}\newcommand{\superstop}{\ensuremath{p^*}\xspace}\newcommand{\sourceStop}{\ensuremath{s}\xspace}\newcommand{\targetStop}{\ensuremath{t}\xspace}\newcommand{\sourceSet}{\ensuremath{\cal S}\xspace}\newcommand{\targetSet}{\ensuremath{\cal T}\xspace}\newcommand{\sourceLocation}{\ensuremath{s^*}\xspace}\newcommand{\targetLocation}{\ensuremath{t^*}\xspace}\newcommand{\event}{\ensuremath{e}}\newcommand{\lab}{\ensuremath{L}}\newcommand{\flabel}{\ensuremath{\lab_f}}\newcommand{\rlabel}{\ensuremath{\lab_b}}\newcommand{\stoplabel}{\ensuremath{SL}}\newcommand{\fslabel}{\ensuremath{\stoplabel_f}}\newcommand{\rslabel}{\ensuremath{\stoplabel_b}}\newcommand{\eventTime}{\ensuremath{\texttt{time}}}\newcommand{\eventStop}{\ensuremath{\texttt{stop}}}\newcommand{\stopSet}{\ensuremath{S}}\newcommand{\tripSet}{\ensuremath{T}}\newcommand{\footSet}{\ensuremath{F}}\newcommand{\eventSet}{\ensuremath{E}}\newcommand{\journey}{\ensuremath{j}}\newcommand{\ajourney}{\ensuremath{\journey_1}\xspace}\newcommand{\bjourney}{\ensuremath{\journey_2}\xspace}\newcommand{\stopEventList}[1]{\ensuremath{E(#1)}}\newcommand{\avertex}{\ensuremath{u}\xspace}\newcommand{\bvertex}{\ensuremath{v}\xspace}\newcommand{\cvertex}{\ensuremath{w}\xspace}\newcommand{\ahub}{\ensuremath{h}\xspace}\newcommand{\atime}{\eventTime\xspace}\DeclareMathOperator{\dist}{dist}\DeclareMathOperator{\mtt}{mtt}\def\comment#1{}\def\withcomments{\newcounter{mycommentcounter}\def\comment##1{\refstepcounter{mycommentcounter}\ifhmode\unskip{\dimen1=\baselineskip \divide\dimen1 by 2 \raise\dimen1\llap{\tiny\bfseries \textcolor{red}{-\themycommentcounter-}}}\fi\marginpar[{\renewcommand{\baselinestretch}{0.8}\hspace*{-1em}\begin{minipage}{10em}\footnotesize [\themycommentcounter] \raggedright ##1\end{minipage}}]{\renewcommand{\baselinestretch}{0.8}\begin{minipage}{10em}\footnotesize [\themycommentcounter]: \raggedright ##1\end{minipage}}}}\newcommand{\email}[1]{\texttt{#1}}\title{Public Transit Labeling\thanks{An extended abstract of this paper has been accepted at the 14th International Symposium on Experimental Algorithms~(SEA'15). Work done mostly while all authors were at Microsoft Research Silicon Valley.}}\author[1]{Daniel Delling}\author[2]{Julian Dibbelt}\author[3]{Thomas Pajor}\author[4]{Renato~F.~Werneck}\affil[1]{Sunnyvale, USA, \email{daniel.delling@gmail.com}}\affil[2]{Karlsruhe Institute of Technology, Germany, \email{dibbelt@kit.edu}}\affil[3]{Microsoft Research, USA, \email{tpajor@microsoft.com}}\affil[4]{San Francisco, USA, \email{rwerneck@acm.org}}\date{May 5, 2015}\begin{document}

\maketitle

\begin{abstract} We study the journey planning problem in public transit networks. Developing efficient preprocessing-based speedup techniques for this problem has been challenging: current approaches either require massive preprocessing effort or provide limited speedups. Leveraging recent advances in Hub Labeling, the fastest algorithm for road networks, we revisit the well-known time-expanded model for public transit. Exploiting domain-specific properties, we provide simple and efficient algorithms for the earliest arrival, profile, and multicriteria problems, with queries that are orders of magnitude faster than the state of the art. \end{abstract}

\section{Introduction} \label{sec:intro} Recent research on route planning in transportation networks~\cite{bdgmpsww-rptn-14} has produced several speedup techniques varying in preprocessing time, space, query performance, and simplicity. Overall, queries on road networks are several orders of magnitude faster than on public transit~\cite{bdgmpsww-rptn-14}. Our aim is to reduce this gap.

There are many natural query types in public transit. An \emph{earliest arrival} query seeks a journey that arrives at a target stop \targetStop as early as possible, given a source stop \sourceStop and a departure time~(\eg, ``now''). A \emph{multicriteria} query also considers the number of transfers when traveling from \sourceStop\ to \targetStop. A \emph{profile} query reports all quickest journeys between two stops within a time range.

These problems can be approached by variants of Dijkstra's algorithm~\cite{d-ntpcg-59} applied to a graph modeling the public transit network, with various techniques to handle time-dependency~\cite{pswz-emtip-08}. In particular, the \emph{time-expanded}~(TE) graph encodes time in the vertices, creating a vertex for every \emph{event} (\eg, a train departure or arrival at a stop at a specific time). Newer approaches, like CSA~\cite{dpsw-isftr-13} and RAPTOR~\cite{dpw-rbptr-14}, work directly on the timetable. Speedup techniques~\cite{bdgmpsww-rptn-14} such as Transfer Patterns~\cite{bceghrv-frvlp-10,bs-fbspt-14}, Timetable Contraction Hierarchies~\cite{g-ctnrt-10}, and ACSA~\cite{sw-csa-13} use preprocessing to create auxiliary data that is then used to accelerate queries.

For aperiodic timetables, the TE model yields a \emph{directed acyclic graph}~(DAG), and several public transit query problems translate to reachability problems. Although these can be solved by simple graph searches, this is too slow for our application. Different methodologies exist to enable faster reachability computation~\cite{chwf-tflat-13,jw-sfsro-13,ms-prcha-14,sabw-ferra-13,yaiy-fsrqg-13,ycz-grail-10,zlwx-rqldg-14}. In particular, the \emph{2-hop labeling}~\cite{chkz-rdqhl-03} scheme associates with each vertex two labels (forward and backward); reachability (or shortest-path distance) can be determined by intersecting the source's forward label and the target's backward label. On continental road networks, 2-hop labeling distance queries take less than a microsecond~\cite{adgw-hhlsp-12}.

In this work, we adapt 2-hop labeling to public transit networks, improving query performance by orders of magnitude over previous methods, while keeping preprocessing time practical. Starting from the time-expanded graph model~(Section~\ref{sec:basic}), we extend the labeling scheme by carefully exploiting properties of public transit networks (Section~\ref{sec:leverage}). Besides earliest arrival and profile queries, we address multicriteria and location-to-location queries, as well as reporting the full journey description quickly~(Section~\ref{sec:practical}). We validate our Public Transit Labeling~(PTL) algorithm by careful experimental evaluation on large metropolitan and national transit networks (Section~\ref{sec:exp}), achieving queries within microseconds.

\section{Preliminaries} \label{sec:prelim}

Let~$G = (V,A)$ be a (weighted) \emph{directed graph}, where~$V$ is the set of vertices and~$A$ the set of arcs. An arc between two vertices $\avertex,\bvertex \in V$ is denoted by~$(\avertex,\bvertex)$. A \emph{path} is a sequence of adjacent vertices. A vertex~\bvertex is \emph{reachable} from a vertex~\avertex if there is a path from~\avertex to \bvertex. A \emph{DAG} is a graph that is both directed and acyclic.

We consider \emph{aperiodic} timetables, consisting of sets of stops~\stopSet, events~\eventSet, trips~\tripSet, and footpaths~\footSet. \emph{Stops} are distinct locations where one can board a transit vehicle~(such as bus stops or subway platforms). \emph{Events} are the scheduled departures and arrivals of vehicles. Each event~$\event \in \eventSet$ has an associated stop~$\eventStop(\event)$ and time~$\eventTime(\event)$. Let $\stopEventList{\astop} = \{\event_0(\astop),\ldots,\event_{k_\astop}(\astop)\}$ be the list (ordered by time) of events at a stop~\astop. We set~$\eventTime(\event_i(\astop)) = -\infty$ for $i < 0$, and~$\eventTime(\event_i(\astop)) = \infty$ for $i > k_\astop$. For simplicity, we may drop the index of an event (as in~$\event(\astop) \in \stopEventList{\astop}$) or its stop (as in $\event \in \eventSet$). A \emph{trip} is a sequence of events served by the same vehicle. A pair of a consecutive departure and arrival events of a trip is a \emph{connection}. \emph{Footpaths} model transfers between nearby stops, each with a predetermined walking duration.

A journey planning algorithm outputs a set of \emph{journeys}. A journey is a sequence of trips (each with a pair of pick-up and drop-off stops) and footpaths in the order of travel. Journeys can be measured according to several criteria, such as arrival time or number of transfers. A journey~\ajourney \emph{dominates} a journey~\bjourney if and only if~\ajourney is no worse in any criterion than~\bjourney. If~$\ajourney$ and~$\bjourney$ are equal in all criteria, we break ties arbitrarily. A set of non-dominated journeys is called a \emph{Pareto set}. Multicriteria Pareto optimization is NP-hard in general, but practical for natural criteria in public transit networks~\cite{dpw-rbptr-14,dpsw-isftr-13,mw-op-06,pswz-emtip-08}. A journey is \emph{tight} if there is no other journey between the same source and target that dominates it in terms of departure and arrival time, \eg, that departs later and arrives earlier.

Given a timetable, stops~$\sourceStop$ and~$\targetStop$, and a departure time~$\depTime$, the~\emph{$(\sourceStop,\targetStop,\depTime)$-earliest arrival}~(EA) problem asks for an~$\sourceStop$--$\targetStop$ journey that arrives at~\targetStop as early as possible and departs at~\sourceStop no earlier than~$\depTime$. The~\emph{$(\sourceStop,\targetStop)$-profile} problem asks for a Pareto set of all tight journeys between~$\sourceStop$ and~$\targetStop$ over the entire timetable period. Finally, the~\emph{$(\sourceStop,\targetStop,\depTime)$-multicriteria}~(MC) problem asks for a Pareto set of journeys departing at~$\sourceStop$ no earlier than~$\depTime$ and minimizing the criteria arrival time and number of transfers. We focus on computing the \emph{values} of the associated optimization criteria of the journeys~(\ie, departure time, arrival times, number of transfers), which is enough for many applications. Section~\ref{sec:practical} discusses how the full journey description can be obtained with little overhead.

Our algorithms are based on the 2-hop~labeling scheme for directed graphs~\cite{chkz-rdqhl-03}. It associates with every vertex~\bvertex a \emph{forward label}~$\flabel(\bvertex)$ and a \emph{backward label}~$\rlabel(\bvertex)$. In a \emph{reachability labeling}, labels are subsets of~$V$, and vertices $\avertex \in \flabel(\bvertex) \cup \rlabel(\bvertex)$ are \emph{hubs} of \bvertex. Every hub in $\flabel(\bvertex)$ must be reachable from $\bvertex$, which in turn must be reachable by every hub in $\rlabel(\bvertex)$. In addition, labels must obey the \emph{cover property}: for any pair of vertices~\avertex and~\bvertex, the intersection~$\flabel(\avertex) \cap \rlabel(\bvertex)$ must contain at least one hub on a~\avertex--\bvertex~path~(if it exists). It follows from this definition that~$\flabel(\avertex) \cap \rlabel(\bvertex) \neq \emptyset$ if and only if~\bvertex is reachable from~\avertex.

In a \emph{shortest path labeling}, each hub~$\avertex \in \flabel(\bvertex)$ also keeps the associated distance~$\dist(\avertex, \bvertex)$, or~$\dist(\bvertex,\avertex)$ for backward labels, and the cover property requires $\flabel(\avertex) \cap \rlabel(\bvertex)$ to contain at least one hub on a \emph{shortest} \mbox{\avertex--\bvertex~path}. If labels are kept sorted by hub ID, a \emph{distance label query} efficiently computes $\dist(\avertex,\bvertex)$ by a coordinated linear sweep over~$\flabel(\avertex)$ and $\rlabel(\bvertex)$, finding the hub~$\cvertex \in \flabel(\avertex) \cap \rlabel(\bvertex)$ that minimizes $\dist(\avertex,\cvertex) + \dist(\cvertex,\bvertex)$. In contrast, a \emph{reachability label query} can stop as soon as any matching hub is found.

In general, smaller labels lead to less space and faster queries. Many algorithms to compute labelings have been proposed~\cite{adgw-hhlsp-12,aiy-f-13,chwf-tflat-13,jw-sfsro-13,yaiy-fsrqg-13,zlwx-rqldg-14}, often for restricted graph classes. We leverage (as a black box) the recent RXL algorithm~\cite{dgpw-rdqmn-14}, which efficiently computes small shortest path labelings for a variety of graph classes at scale. It is a sampling-based greedy algorithm that builds labels one hub at a time, with priority to vertices that cover as many relevant paths as possible.

Different approaches for transforming a timetable into a graph exist~(see~\cite{pswz-emtip-08} for an overview). In this work, we focus on the \emph{time-expanded model}. Since it uses scalar arc costs, it is a natural choice for adapting the labeling approach. In contrast, the \emph{time-dependent model}~(another popular approach) associates functions with the arcs, which makes adaption more difficult.

\section{Basic Approach} \label{sec:basic}

We build the time-expanded graph from the timetable as follows. We group all departure and arrival events by the stop where they occur. We sort all events at a stop by time, merging events that happen at the same stop and time. We then add a vertex for each unique event, a \emph{waiting arc} between two consecutive events of the same stop, and a \emph{connection arc} for each connection (between the corresponding departure and arrival event). The cost of arc $(\avertex,\bvertex)$ is $\eventTime(\bvertex) - \eventTime(\avertex)$, \ie, the time difference of the corresponding events. To account for footpaths between two stops $a$ and $b$, we add, from each vertex at stop $a$, a \emph{foot arc} to the first reachable vertex at~$b$ (based on walking time), and vice versa. As events and vertices are tightly coupled in this model, we use the terms interchangeably.

Any label generation scheme~(we use RXL~\cite{dgpw-rdqmn-14}) on the time-expanded graph creates two~(forward and backward) \emph{event labels} for every vertex~(event), enabling \emph{event-to-event queries}. For our application \emph{reachability} labels~\cite{yaiy-fsrqg-13}, which only store hubs~(without distances), suffice. First, since all arcs point to the future, time-expanded graphs are DAGs. Second, if an event \event\ is reachable from another event $\event'$ (\ie, $\flabel(\event') \cap \rlabel(\event) \neq \emptyset$), we can compute the time to get from $\event'$\ to \event\ as $\eventTime(\event) - \eventTime(\event')$. In fact, \emph{all} paths between two events have equal cost.

In practice, however, event-to-event queries are of limited use, as they require users to specify both departure \emph{and} arrival times, one of which is usually unknown. Therefore, we discuss earliest arrival and profile queries, which \emph{optimize} arrival time and are thus more meaningful. See Section~\ref{sec:practical} for multicriteria queries.

\paragraph{Earliest Arrival Queries.} Given event labels, we answer an ($\sourceStop,\targetStop,\depTime$)-EA query as follows. We first find the earliest event $\event_i(\sourceStop) \in \stopEventList{\sourceStop}$ at the source stop~\sourceStop that suits the departure time, \ie, with $\eventTime(\event_i(\sourceStop)) \geq \depTime$ and \mbox{$\eventTime(\event_{i-1}(\sourceStop)) < \depTime$}. Next, we search at the target stop~\targetStop for the earliest event $\event_j(\targetStop) \in \stopEventList{\targetStop}$ that is reachable from~$\event_i(\sourceStop)$ by testing whether~$\flabel(\event_i(\sourceStop)) \cap \rlabel(\event_j(\targetStop)) \neq \emptyset$ and \mbox{$\flabel(\event_i(\sourceStop)) \cap \rlabel(\event_{j-1}(\targetStop)) = \emptyset$}. Then, $\eventTime(\event_j(\targetStop))$ is the earliest arrival time. One could find $\event_j(\targetStop)$ using linear search (which is simple and cache-friendly), but binary search is faster in theory and in practice. To accelerate queries, we \emph{prune} (skip) all events~$\event(\targetStop)$ with~$\eventTime(\event(\targetStop)) < \depTime$, since~$\flabel(\event_i(\sourceStop)) \cap \rlabel(\event(\targetStop)) = \emptyset$ always holds in such cases. Moreover, to avoid evaluating $\flabel(\event_i(\sourceStop))$ multiple times, we use \emph{hash-based queries}~\cite{dgpw-rdqmn-14}: we first build a hash set of the hubs in $\flabel(\event_i(\sourceStop))$, then check the reachability for an event $\event(\targetStop)$ by probing the hash with hubs $h \in \rlabel(\event(\targetStop))$.

\paragraph{Profile Queries.} To answer an ($\sourceStop,\targetStop$)-profile query, we perform a coordinated sweep over the events at $\sourceStop$ and $\targetStop$. For the current event~$\event_i(\sourceStop) \in \stopEventList{\sourceStop}$ at the source stop~(initialized to the earliest event~$\event_0(\sourceStop) \in \stopEventList{\sourceStop}$), we find the first event~\mbox{$\event_j(\targetStop) \in \stopEventList{\targetStop}$} at the target stop that is reachable, \ie, such that \mbox{$\flabel(\event_i(\sourceStop)) \cap \rlabel(\event_j(\targetStop)) \neq \emptyset$} and~$\flabel(\event_i(\sourceStop)) \cap \rlabel(\event_{j-1}(\targetStop)) = \emptyset$. This gives us the earliest arrival time~$\eventTime(\event_j(\targetStop))$. To identify the latest departure time from \sourceStop\ for that earliest arrival event (and thus have a tight journey), we increase $i$ until~$\flabel(\event_{i}(\sourceStop)) \cap \rlabel(\event_j(\targetStop)) = \emptyset$, then add~$(\eventTime(\event_{i-1}(\sourceStop)),\eventTime(\event_j(\targetStop)))$ to the profile. We repeat the process starting from the events~$\event_{i}(\sourceStop)$ and~$\event_{j+1}(\targetStop)$. Since we increase either~$i$ or~$j$ after each intersection test, the worst-case time to find all tight journeys is linear in the number of events~(at~$\sourceStop$ and~$\targetStop$) multiplied by the size of their largest label.

\section{Leveraging Public Transit} \label{sec:leverage} Our approach can be refined to exploit features specific to public transit networks. As described so far, our labeling scheme maintains reachability information for \emph{all pairs} of events (by covering all paths of the time-expanded graph, breaking ties arbitrarily). However, in public transit networks we actually are only interested in \emph{certain paths}. In particular, the labeling does \emph{not} need to cover any path ending at a departure event (or beginning at an arrival event). We can thus discard forward labels from arrival events and backward labels from departure events.

\paragraph{Trimmed Event Labels.} Moreover, we can disregard paths representing dominated journeys that depart earlier and arrive later than others (\ie, journeys that are not tight, \cf~Section~\ref{sec:prelim}). Consider all departure events of a stop. If a certain hub is reachable from event~$\event_i(\sourceStop)$, then it is also reachable from $\event_0(\sourceStop), \ldots, \event_{i-1}(\sourceStop)$, and is thus potentially added to the forward labels of all these earlier events. In fact, experiments show that on average the same hub is added to 1.8--5.0 events per stop~(depending on the network). We therefore compute \emph{trimmed event labels} by discarding all but the latest occurrence of each hub from the forward labels. Similarly, we only keep the earliest occurrence of each hub in the backward labels. (Preliminary experiments have shown that we obtain very similar label sizes with a much slower algorithm that greedily covers tight journeys explicitly~\cite{adgw-hhlsp-12,dgpw-rdqmn-14}.)

Unfortunately, we can no longer just apply the query algorithms from Section~\ref{sec:basic} with trimmed event labels: if the selected departure event at~$\sourceStop$ does not correspond to a tight journey toward~$\targetStop$, the algorithm will not find a solution~(though one might exist). One could circumvent this issue by also running the algorithm from subsequent departure events at~$\sourceStop$, which however may lead to quadratic query complexity in the worst case~(for both EA and profile queries).

\paragraph{Stop Labels.} We solve this problem by working with \emph{stop labels}: For each stop~\astop, we merge all forward event labels $\flabel(\event_0(\astop)), \ldots, \flabel(\event_k(\astop))$ into a forward stop label \fslabel(\astop), and all backward event labels into a backward stop label \rslabel(\astop). Similar to distance labels, each stop label~$\stoplabel(\astop)$ is a list of pairs~$(\ahub, \atime_\astop(\ahub))$, each containing a hub and a time, sorted by hub. For a forward label, $\atime_\astop(\ahub)$ encodes the latest departure time from~\astop to reach hub~\ahub. More precisely, let $h$ be a hub in an event label~$\flabel(\event_i(\astop))$: we add the pair~$(\ahub, \eventTime(\event_i(\astop)))$ to the stop label \fslabel(\astop) only if $\ahub \notin \flabel(\event_j(\astop)), j>i$, \ie, only if \ahub does not appear in the label of another event with a later departure time at the stop. Analogously, for backward stop labels, $\atime_\astop(\ahub)$ encodes the earliest arrival time at $p$ from~$h$.

By restricting ourselves to these entries, we effectively discard dominated~(non-tight) journeys to these hubs. It is easy to see that these stop labels obey a \emph{tight journey cover property}: for each pair of stops \sourceStop and \targetStop, $\fslabel(\sourceStop) \cap \rslabel(\targetStop)$ contains at least one hub on each tight journey between them (or any equivalent journey that departs and arrives at the same time; recall from Section~\ref{sec:prelim} that we allow arbitrary tie-breaking). This property does \emph{not}, however, imply that the label intersection \emph{only} contains tight journeys: for example, $\fslabel(\sourceStop)$ and $\rslabel(\targetStop)$ could share a hub that is important for long distance travel, but not to get from~\sourceStop~to~\targetStop. The remainder of this section discusses how we handle this fact during queries.

\paragraph{Stop Label Profile Queries.} To run an (\sourceStop,\targetStop)-profile query on stop labels, we perform a coordinated sweep over both labels~\fslabel(\sourceStop) and \rslabel(\targetStop). For every matching hub~$\ahub$, \ie, $(\ahub, \atime_\sourceStop(\ahub)) \in \fslabel(\sourceStop)$ and $(\ahub, \atime_\targetStop(\ahub)) \in \rslabel(\targetStop)$, we consider the journey induced by~$(\atime_\sourceStop(\ahub), \atime_\targetStop(\ahub))$ for output. However, since we are only interested in reporting tight journeys, we maintain~(during the algorithm) a tentative set of tight journeys, removing dominated journeys from it on-the-fly. (We found this to be faster than adding all journeys during the sweep and only discarding dominated journeys at the end.) We can further improve the efficiency of this approach in practice by (globally) reassigning hub IDs by the time of day. Note that every hub~$\ahub$ of a stop label is still also an event and carries an event time~$\eventTime(\ahub)$. (Not to be confused with $\atime_\sourceStop(\ahub)$ and $\atime_\targetStop(\ahub)$.) We assign sequential IDs to all hubs~$\ahub$ in order of increasing $\eventTime(\ahub)$, thus ensuring that hubs in the label intersection are enumerated chronologically. Note that this does not imply that journeys are enumerated in order of departure or arrival time, since each hub $\ahub$ may appear anywhere along its associated journey. However, preliminary experiments have shown that this approach leads to fewer insertions into the tentative set of tight journeys, reducing query time. Moreover, as in shortest path labels~\cite{dgpw-rdqmn-14}, we improve cache efficiency by storing the values for hubs and times separately in a stop label, accessing times only for matching hubs.

Overall, stop and event labels have different trade-offs: maintaining the profile requires less effort with event labels~(any discovered journey is already tight), but fewer hubs are scanned with stop labels~(there are no duplicate hubs).

\paragraph{Stop Label Earliest Arrival Queries.} Reassigned hub IDs also enable fast $(\sourceStop,\targetStop,\depTime)$-EA queries. We use binary search in $\fslabel(\sourceStop)$ and $\rslabel(\targetStop)$ to find the earliest relevant hub~\ahub, \ie, with $\eventTime(\ahub) \geq \depTime$. From there, we perform a linear coordinated sweep as in the profile query, finding $(\ahub, \atime_\sourceStop(\ahub)) \in \fslabel(\sourceStop)$ and $(\ahub, \atime_\targetStop(\ahub)) \in \rslabel(\targetStop)$. However, instead of maintaining tentative profile entries~$(\atime_\sourceStop(\ahub), \atime_\targetStop(\ahub))$, we ignore solutions that depart too early~(\ie, $\atime_\sourceStop(\ahub) < \depTime$), while picking the hub~$\ahub^*$ that minimizes the tentative best arrival time~$\atime_\targetStop(\ahub^*)$. (Note that $\eventTime(\ahub) \geq \depTime$ does not imply $\atime_\sourceStop(\ahub) \geq \depTime$.) Once we scan a hub~$h$ with~\mbox{$\eventTime(\ahub) \geq \atime_\targetStop(\ahub^*)$}, the tentative best arrival time cannot be improved anymore, and we stop the query. For practical performance, \emph{pruning} the scan, so that we only sweep hubs~\ahub between $\depTime \leq \atime(\ahub) \leq \atime_\targetStop(\ahub^*)$, is very important.

\section{Practical Extensions} \label{sec:practical}

So far, we presented stop-to-stop queries, which report the departure and arrival times of the quickest journey(s). In this section, we address multicriteria queries, general location-to-location requests, and obtaining detailed journey descriptions.

\paragraph{Multicriteria Optimization and Minimum Transfer Time.} Besides optimizing arrival time, many users also prefer journeys with fewer transfers. To solve the underlying multicriteria optimization problem, we adapt our labeling approach by (1)~encoding transfers as arc costs in the graph, (2)~computing shortest path labels based on these costs (instead of reachability labels on an unweighted graph), and (3)~adjusting the query algorithm to find the Pareto set of solutions.

Reconsider the earliest arrival graph from Section~\ref{sec:basic}. As before, we add a vertex for each unique event, linking consecutive events at the same stop with waiting arcs of cost~0. However, each connection arc~$(u,w)$ in the graph is subdivided by an intermediate \emph{connection vertex}~$v$, setting the cost of arc~$(u,v)$ to 0 and the cost of arc~$(v,w)$ to 1. By interpreting costs of 1 as leaving a vehicle, we can count the number of trips taken along any path. To model staying in the vehicle, consecutive connection vertices of the same trip are linked by zero-cost arcs.

A shortest path labeling on this graph now encodes the number of transfers as the shortest path distance between two events, while the duration of the journey can still be deduced from the time difference of the events. Consider a fixed source event~$\event(\sourceStop)$ and the arrival events of a target stop~$\event_0(\targetStop), \event_1(\targetStop), \ldots$ in order of increasing time. The minimum number of transfers required to reach the target stop~$\targetStop$ never increases with arrival times. (Hence, the whole Pareto set~$P$ of multicriteria solutions can be computed with a single Dijkstra run~\cite{pswz-emtip-08}.)

We exploit this property to compute~$(\sourceStop,\targetStop,\depTime)$-EA multicriteria~(MC) queries from the labels as follows. We initialize~$P$ as the empty set. We then perform an~$(\sourceStop,\targetStop,\depTime)$-EA query~(with all optimizations described in Section~\ref{sec:basic}) to compute the \emph{fastest} journey in the solution, \ie, the one with most transfers. We add this journey to~$P$. We then check~(by performing distance label queries) for each subsequent event at~$\targetStop$ whether there is a journey with fewer transfers~(than the most recently added entry of~$P$), in which case we add the journey to~$P$ and repeat. The MC~query ends once the last event at the target stop has been processed. We can stop earlier with the following optimization: we first run a distance label query on the \emph{last} event at~\targetStop\ to obtain the \emph{smallest} possible number of transfers to travel from~$\sourceStop$ to~$\targetStop$. We may then already stop the MC~query once we add a journey to~$P$ with this many transfers. Note that, since we do not need to check for domination in~$P$ explicitly, our algorithm maintains~$P$ in constant time per added journey.

\paragraph{Minimum Transfer Times.} Transit agencies often model an entire station with multiple platforms as a single stop and account for the time required to change trips inside the station by associating a \emph{minimum transfer time}~$\mtt(\astop)$ with each stop~$\astop$. To incorporate them into the EA graph, we first locally replace each affected stop~$\astop$ by a \emph{set} of new stops~$\superstop$, distributing \emph{conflicting} trips~(between which transferring is impossible due to~$\mtt(\astop)$) to different stops of~$\superstop$. We then add footpaths between all pairs of stops in~$\superstop$ with length~$\mtt(\astop)$. A small set~$\superstop$ can be computed by solving an appropriate coloring problem~\cite{dkp-pcbcp-12}. For the MC graph, we need not change the input. Instead, it is sufficient to \emph{shift} each arrival event~$\event \in \stopEventList{\astop}$ by adding~$\mtt(\astop)$ to~$\eventTime(\event)$ before creating the vertices.

\paragraph{Location-to-Location Queries.}

A query between arbitrary locations~$\sourceLocation$ and~$\targetLocation$, which may employ walking or driving as the first and last legs of the journey, can be handled by a two-stage approach. It first computes sets~$\sourceSet$ and~$\targetSet$ of relevant stops near the origin~$\sourceLocation$ and destination~$\targetLocation$ that can be reached by car or on foot. With that information, a \emph{forward superlabel}~\cite{adfgw-hldbl-12} is built from all forward stop labels associated with \sourceSet. For each entry $(\ahub, \atime_\astop(\ahub)) \in \fslabel(\astop)$ in the label of stop~$\astop \in \sourceSet$, we adjust the departure time~$\atime_\sourceLocation(\ahub) = \atime_\astop(\ahub) - \dist(\sourceLocation, \astop)$ so that the journey starts at \sourceLocation and add $(\ahub, \atime_\sourceLocation(\ahub))$ to the superlabel. For duplicate hubs that occur in multiple stop labels, we keep only the latest departure time from \sourceLocation. This can be achieved with a coordinated sweep, always adding the next hub of minimum ID. A \emph{backward superlabel} (for $\targetSet$) is built analogously. For location-to-location queries, we then simply run our stop-label-based EA and profile query algorithms using the superlabels. In practice, we need not build superlabels explicitly but can simulate the building sweep during the query~(which in itself is a coordinated sweep over two labels). A similar approach is possible for event labels. Moreover, point-of-interest queries~(such as finding the closest restaurants to a given location) can be computed by applying known techniques~\cite{adfgw-hldbl-12} to these superlabels.

\paragraph{Journey Descriptions.} While for many applications it suffices to report departure and arrival times (and possibly the number of transfers) per journey, sometimes a more detailed description is needed. We could apply known path unpacking techniques~\cite{adfgw-hldbl-12} to retrieve the full sequence of connections~(and transfers), but in public transit it is usually enough to report the list of trips with associated transfer stops. We can accomplish that by storing with each hub the sequences of trips~(and transfer stops) for travel between the hub and its label vertex.

\section{Experiments} \label{sec:exp}

\paragraph{Setup.}

We implemented all algorithms in C++ using Visual Studio 2013 with full optimization. All experiments were conducted on a machine with two 8-core Intel Xeon E5-2690 CPUs and 384\,GiB of DDR3-1066 RAM, running Windows 2008R2 Server. All runs are \emph{sequential}. We use at most 32 bits for distances.

\begin{table} \centering \caption{\label{tab:sizes}Size of timetables and the earliest arrival~(EA) and multicriteria~(MC) graphs.} \setlength{\tabcolsep}{0.9ex} \begin{tabular}{@{}lrrrrrrrrr@{}} \toprule &&&&&& \multicolumn{2}{c}{EA Graph} & \multicolumn{2}{c}{MC Graph} \\ \cmidrule(lr){7-8}\cmidrule(l){9-10} Instance & Stops & Conns & Trips & Footp. & Dy. & $|V|$ & $|A|$ & $|V|$ & $|A|$ \\ \midrule London&20.8\,k&5,133\,k&133\,k&45.7\,k&1&4,719\,k&51,043\,k&9,852\,k&72,162\,k\\ Madrid&4.7\,k&4,527\,k&165\,k&1.3\,k&1&3,003\,k&13,730\,k&7,530\,k&34,505\,k\\ Sweden&51.1\,k&12,657\,k&548\,k&1.1\,k&2&8,151\,k&34,806\,k&20,808\,k&93,194\,k\\ Switzerland&27.1\,k&23,706\,k&2,198\,k&29.8\,k&2&7,979\,k&49,656\,k&31,685\,k&170,503\,k\\ \bottomrule \end{tabular} \end{table} We consider four realistic inputs: the metropolitan networks of London (\url{data.london.gov.uk}) and Madrid (\url{emtmadrid.es}), and the national networks of Sweden (\url{trafiklab.se}) and Switzerland (\url{gtfs.geops.ch}). London includes all modes of transport, Madrid contains only buses, and the national networks contain both long-distance and local transit. We consider 24-hour timetables for the metropolitan networks, and two days for national ones (to enable overnight journeys). Footpaths were generated using a known heuristic~\cite{dkp-pcbcp-12} for Madrid; they are part of the input for the other networks. See~\tablename~\ref{tab:sizes} for size figures of the timetables and resulting graphs. The average number of unique events per stop ranges from 160 for Sweden to 644 for Madrid. (Recall from Section~\ref{sec:basic} that we merge all coincident events at a stop.) Note that no two instances dominate each other~(\wrt number of stops, connections, trips, events per stop, and footpaths).

\paragraph{Preprocessing.}

\begin{table}[b] \setlength{\tabcolsep}{0.9ex} \centering \caption{Preprocessing figures. Label sizes are averages of forward and backward labels.} \label{tab:prepro} \begin{tabular}{@{}lrrrrrrrrrr@{}} \toprule & \multicolumn{6}{c}{\textbf{Earliest Arrival}} & \multicolumn{4}{c}{\textbf{Multicriteria}}\\ \cmidrule(lr){2-7}\cmidrule(l){8-11} & & \multicolumn{3}{c}{Event Labels} & \multicolumn{2}{c}{Stop Labels} & & \multicolumn{3}{c}{Event Labels} \\ \cmidrule(lr){3-5}\cmidrule(lr){6-7}\cmidrule(l){9-11} & RXL & Hubs & Hubs & Space & Hubs & Space & RXL & Hubs & Hubs & Space \\ Instance & [h:m] & p.\,lbl & p.\,stop & [MiB] & p.\,stop & [MiB] & [h:m] & p.\,lbl & p.\,stop & [MiB]\\ \midrule London&0:54&70&15,480&1,334&7,075&1,257&49:19&734&162,565&26,871\\ Madrid&0:25&77&49,247&963&9,830&403&10:55&404&258,008&10,155\\ Sweden&0:32&37&5,630&1,226&1,536&700&36:14&190&29,046&12,637\\ Switzerland&0:42&42&11,189&1,282&2,970&708&61:36&216&58,022&12,983\\ \bottomrule \end{tabular} \end{table}

\tablename~\ref{tab:prepro} reports preprocessing figures for the unweighted earliest arrival graph~(which also enables profile queries) and the multicriteria graph. For earliest arrival~(EA), preprocessing takes well below an hour and generates about one gigabyte, which is quite practical. Although there are only 37--70 hubs per label, the total number of hubs per stop (\ie, the combined size of all labels) is quite large~(5,630--49,247). By eliminating redundancy (\cf~Section~\ref{sec:leverage}), stop labels have only a fifth as many hubs (for Madrid). Even though they need to store an additional distance value per hub, total space usage is still smaller. In general, \emph{average} labels sizes (though not total space) are higher for metropolitan instances. This correlates with the higher number of daily journeys in these networks.

Preprocessing the multicriteria~(MC) graph is much more expensive: times increase by a factor of~26.2--54.8 for the metropolitan and~67.9--88 for the national networks. On Madrid, Sweden, and Switzerland labels are five times larger compared to EA, and on London the factor is even more than ten. This is immediately reflected in the space consumption, which is up to 26\,GiB~(London).

\paragraph{Queries.}

We now evaluate query performance. For each algorithm, we ran~100,000 queries between random source and target stops, at random departure times between~0:00 and~23:59~(of the first day). \tablename~\ref{tab:ea-queries} reports detailed figures, organized in three blocks: event label EA queries, stop label EA queries, and profile queries (with both event and stop labels). We discuss MC queries later.

We observe that event labels result in extremely fast EA queries (6.9--14.7\,\musec), even without optimizations. As expected, pruning and hashing reduce the number of accesses to labels and hubs~(see columns ``Lbls.'' and ``Hubs''). Although binary search cannot stop as soon as a matching hub is found~(see the ``='' column), it accesses fewer labels and hubs, achieving query times below~3\,\musec\ on all instances.

Using stop labels~(\cf~Section~\ref{sec:leverage}) in their basic form is significantly slower than using event labels. With pruning enabled, however, query times (3.6--6.2\,\musec) are within a factor of two of the event labels, while saving a factor of 1.1--2.4 in space. For profile queries, stop labels are clearly the best approach. It scans up to a factor of 5.1 fewer hubs and is up to 3.3 times faster, computing the profile of the full timetable period in under 80\,\musec~on all instances. The difference in factors is due to the overhead of maintaining the Pareto set during the stop label query.

\begin{table} \setlength{\tabcolsep}{0.7ex} \centering \caption{Evaluating earliest arrival queries. Bullets~($\enabled$) indicate different features: profile query~(Prof.), stop labels~(St.\,lbs.), pruning~(Prn.), hashing~(Hash), and binary search~(Bin.). The column ``='' indicates the average number of matched hubs.} \label{tab:ea-queries} \begin{tabular}{@{}cccccrrrrrrrrrrrr@{}} \toprule &&&&&\multicolumn{4}{c}{\textbf{London}}&\multicolumn{4}{c}{\textbf{Sweden}}&\multicolumn{4}{c}{\textbf{Switzerland}}\\ \cmidrule(lr){6-9}\cmidrule(lr){10-13}\cmidrule(lr){14-17} \feature{Prof.} & \feature{St.\,lbs.} & \feature{Prn.} & \feature{Hash} & \feature{Bin.} & Lbls. & Hubs & = & [\textmu{}s] & Lbls. & Hubs & = & [\textmu{}s] & Lbls. & Hubs & = & [\textmu{}s] \\ \midrule \disabled&\disabled&\disabled&\disabled&\disabled&108.4&6,936&1&14.7&68.0&2,415&1&6.9&89.0&3,485&1&8.7\\ \disabled&\disabled&\enabled&\disabled&\disabled&16.1&1,360&1&5.9&34.4&1,581&1&5.4&33.5&1,676&1&5.8\\ \disabled&\disabled&\enabled&\enabled&\disabled&16.1&1,047&1&4.2&34.4&1,083&1&3.6&33.5&1,151&1&3.8\\ \disabled&\disabled&\enabled&\enabled&\enabled&7.0&332&4&2.8&6.5&179&3&2.1&7.6&204&4&2.1\\ \addlinespace \disabled&\enabled&\disabled&\disabled&\disabled&2.0&13,037&1,126&54.8&2.0&2,855&81&10.0&2.0&5,707&218&20.4\\ \disabled&\enabled&\enabled&\disabled&\disabled&2.0&861&62&6.2&2.0&711&16&3.6&2.0&699&19&3.8\\ \addlinespace \enabled&\disabled&\disabled&\disabled&\disabled&658.5&40,892&211&141.7&423.7&13,590&118&39.4&786.6&29,381&240&81.4\\ \enabled&\enabled&\disabled&\disabled&\disabled&2.0&13,037&1,126&74.3&2.0&2,855&81&12.1&2.0&5,707&218&24.5\\ \bottomrule \end{tabular} \end{table}

\paragraph{Comparison.}

\begin{table} \setlength{\tabcolsep}{1ex} \centering \caption{Comparison with the state of the art. Presentation largely based on~\cite{bdgmpsww-rptn-14}, with some additional results taken from~\cite{bs-fbspt-14}. The first block of techniques considers the EA problem, the second the MC problem and the third the profile problem.} \label{tab:comparison} \begin{tabular}{@{}llrrrcccrrr@{}} \toprule &\multicolumn{4}{c}{\textbf{Instance}}&\multicolumn{3}{c}{\textbf{Criteria}}\\ \cmidrule(lr){2-5}\cmidrule{6-8} &&Stops&Conns&&&&&Prep.&&Query\\ Algorithm&Name&[$\cdot 10^3$]&[$\cdot 10^6$]&Dy.&\feature{Arr.}&\feature{Tran.}&\feature{Prof.}&[h]&Jn.&[ms]\\ \midrule CSA~\cite{dpsw-isftr-13}&London&20.8&4.9&1&\enabled&\disabled&\disabled&---&n/a&1.8\phantom{000}\\ ACSA~\cite{sw-csa-13}&Germany&252.4&46.2&2&\enabled&\disabled&\disabled&0.2&n/a&8.7\phantom{000}\\ CH~\cite{g-ctnrt-10}&Europe~(LD)&30.5&1.7&p&\enabled&\disabled&\disabled&$<$\,0.1&n/a&0.3\phantom{000}\\ TP~\cite{bdgmpsww-rptn-14}&Madrid&4.6&4.8&1&\enabled&\disabled&\disabled&19\phantom{.0}&n/a&0.7\phantom{000}\\ TP~\cite{bs-fbspt-14}&Germany&248.4&13.9&1&\enabled&\disabled&\disabled&249\phantom{.0}&0.9&0.2\phantom{000}\\ PTL&London&20.8&5.1&1&\enabled&\disabled&\disabled&0.9&0.9&0.0028\\ PTL&Madrid&4.7&4.5&1&\enabled&\disabled&\disabled&0.4&0.9&0.0030\\ PTL&Sweden&51.1&12.7&2&\enabled&\disabled&\disabled&0.5&1.0&0.0021\\ PTL&Switzerland&27.1&23.7&2&\enabled&\disabled&\disabled&0.7&1.0&0.0021\\ \addlinespace RAPTOR~\cite{dpw-rbptr-14}&London&20.8&5.1&1&\enabled&\enabled&\disabled&---&1.8&5.4\phantom{000}\\ TP~\cite{bdgmpsww-rptn-14}&Madrid&4.6&4.8&1&\enabled&\enabled&\disabled&185\phantom{.0}&n/a&3.1\phantom{000}\\ TP~\cite{bs-fbspt-14}&Germany&248.4&13.9&1&\enabled&\enabled&\disabled&372\phantom{.0}&1.9&0.3\phantom{000}\\ PTL&London&20.8&5.1&1&\enabled&\enabled&\disabled&49.3&1.8&0.0266\\ PTL&Madrid&4.7&4.5&1&\enabled&\enabled&\disabled&10.9&1.9&0.0643\\ PTL&Sweden&51.1&12.7&2&\enabled&\enabled&\disabled&36.2&1.7&0.0276\\ PTL&Switzerland&27.1&23.7&2&\enabled&\enabled&\disabled&61.6&1.7&0.0217\\ \addlinespace CSA~\cite{dpsw-isftr-13}&London&20.8&4.9&1&\enabled&\disabled&\enabled&---&98.2&161.0\phantom{000}\\ ACSA~\cite{sw-csa-13}&Germany&252.4&46.2&2&\enabled&\disabled&\enabled&0.2&n/a&171.0\phantom{000}\\ CH~\cite{g-ctnrt-10}&Europe~(LD)&30.5&1.7&p&\enabled&\disabled&\enabled&$<$\,0.1&n/a&3.7\phantom{000}\\ TP~\cite{bs-fbspt-14}&Germany&248.4&13.9&1&\enabled&\disabled&\enabled&249\phantom{.0}&16.4&3.3\phantom{000}\\ PTL&London&20.8&5.1&1&\enabled&\disabled&\enabled&0.9&81.0&0.0743\\ PTL&Madrid&4.7&4.5&1&\enabled&\disabled&\enabled&0.4&110.7&0.1119\\ PTL&Sweden&51.1&12.7&2&\enabled&\disabled&\enabled&0.5&12.7&0.0121\\ PTL&Switzerland&27.1&23.7&2&\enabled&\disabled&\enabled&0.7&31.5&0.0245\\ \bottomrule \end{tabular} \end{table}

\tablename~\ref{tab:comparison} compares our new algorithm (indicated as \emph{PTL}, for Public Transit Labeling) to the state of the art and also evaluates multicriteria queries. In this experiment, PTL uses event labels with pruning, hashing and binary search for earliest arrival~(and multicriteria) queries, and stop labels for profile queries. We compare PTL to~CSA~\cite{dpsw-isftr-13} and RAPTOR~\cite{dpw-rbptr-14}~(currently the fastest algorithms without preprocessing), as well as Accelerated CSA~(ACSA)~\cite{sw-csa-13}, Timetable Contraction Hierarchies~(CH)~\cite{g-ctnrt-10}, and Transfer Patterns~(TP)~\cite{bceghrv-frvlp-10,bs-fbspt-14}~(which make use of preprocessing). Since RAPTOR always optimizes transfers~(by design), we only include it for the MC problem. Note that the following evaluation should be taken with a grain of salt, as no standardized benchmark instances exist, and many data sets used in the literature are proprietary. Although precise numbers are not available for several competing methods, it is safe to say they use less space than PTL, particularly for the MC problem.

\tablename~\ref{tab:comparison} shows that PTL queries are very efficient. Remarkably, they are faster on the national networks than on the metropolitan ones: the latter are smaller in most aspects, but have more frequent journeys (that must be covered). Compared to other methods, PTL is~\mbox{2--3}~orders of magnitude faster on London than CSA and RAPTOR for EA~(factor~643), profile~(factor~2,167), and MC~(factor~203) queries. We note, however, that PTL is a point-to-point algorithm~(as are ACSA, TP, and CH); for one-to-all queries, CSA and RAPTOR would be faster.

PTL has 1--2~orders of magnitude faster preprocessing and queries than TP for the EA and profile problems. On Madrid, EA queries are 233~times faster while preprocessing is faster by a factor of~48. Note that Sweden~(PTL) and Germany~(TP) have a similar number of connections, but PTL queries are~95~times faster. (Germany does have more stops, but recall that PTL query performance depends more on the frequency of trips.) For the MC problem, the difference is smaller, but both preprocessing and queries of PTL are still an order of magnitude faster than TP~(up to 48~times for MC queries on Madrid).

Compared to ACSA and CH~(for which figures are only available for the EA and profile problems), PTL has slower preprocessing but significantly faster queries~(even when accounting for different network sizes).

\section{Conclusion} \label{sec:conclusion}

We introduced PTL, a new preprocessing-based algorithm for journey planning in public transit networks, by revisiting the time-expanded model and adapting the Hub Labeling approach to it. By further exploiting structural properties specific to timetables, we obtained simple and efficient algorithms that outperform the current state of the art on large metropolitan and country-sized networks by orders of magnitude for various realistic query types. Future work includes developing tailored algorithms for hub computation~(instead of using RXL as a black box), compressing the labels~(\eg, using techniques from~\cite{bs-fbspt-14} and~\cite{dgpw-rdqmn-14}), exploring other hub representations~(\eg, using trips instead of events, as in 3-hop labeling~\cite{yaiy-fsrqg-13}), using multicore- and instruction-based parallelism for preprocessing and queries, and handling dynamic scenarios~(\eg, temporary station closures and train delays or cancellations~\cite{bdgmpsww-rptn-14}).

\bibliographystyle{plain} \begin{small} 

\begin{thebibliography}{10}

\bibitem{adfgw-hldbl-12} Ittai Abraham, Daniel Delling, Amos Fiat, Andrew~V. Goldberg, and Renato~F. Werneck. \newblock {HLDB}: Location-based services in databases. \newblock In {\em Proceedings of the 20th ACM SIGSPATIAL International Symposium on Advances in Geographic Information Systems (GIS'12)}, pages 339--348. ACM Press, 2012. \newblock Best Paper Award.

\bibitem{adgw-hhlsp-12} Ittai Abraham, Daniel Delling, Andrew~V. Goldberg, and Renato~F. Werneck. \newblock Hierarchical hub labelings for shortest paths. \newblock In {\em Proceedings of the 20th Annual European Symposium on Algorithms (ESA'12)}, volume 7501 of {\em Lecture Notes in Computer Science}, pages 24--35. Springer, 2012.

\bibitem{aiy-f-13} Takuya Akiba, Yoichi Iwata, and Yuichi Yoshida. \newblock Fast exact shortest-path distance queries on large networks by pruned landmark labeling. \newblock In {\em Proceedings of the 2013 ACM SIGMOD International Conference on Management of Data (SIGMOD'13)}, pages 349--360. ACM Press, 2013.

\bibitem{bceghrv-frvlp-10} Hannah Bast, Erik Carlsson, Arno Eigenwillig, Robert Geisberger, Chris Harrelson, Veselin Raychev, and Fabien Viger. \newblock Fast routing in very large public transportation networks using transfer patterns. \newblock In {\em Proceedings of the 18th Annual European Symposium on Algorithms (ESA'10)}, volume 6346 of {\em Lecture Notes in Computer Science}, pages 290--301. Springer, 2010.

\bibitem{bdgmpsww-rptn-14} Hannah Bast, Daniel Delling, Andrew~V. Goldberg, Matthias {M{\"u}ller--Hannemann}, Thomas Pajor, Peter Sanders, Dorothea Wagner, and Renato~F. Werneck. \newblock Route planning in transportation networks. \newblock Technical Report MSR-TR-2014-4, Microsoft Research, 2014.

\bibitem{bs-fbspt-14} Hannah Bast and Sabine Storandt. \newblock Frequency-based search for public transit. \newblock In {\em Proceedings of the 22nd ACM SIGSPATIAL International Conference on Advances in Geographic Information Systems}. ACM Press, November 2014.

\bibitem{chwf-tflat-13} James Cheng, Silu Huang, Huanhuan Wu, and Ada Wai-Chee Fu. \newblock Tf-label: A topological-folding labeling scheme for reachability querying in a large graph. \newblock In {\em Proceedings of the 2013 ACM SIGMOD International Conference on Management of Data (SIGMOD'13)}, pages 193--204. ACM Press, 2013.

\bibitem{chkz-rdqhl-03} Edith Cohen, Eran Halperin, Haim Kaplan, and Uri Zwick. \newblock Reachability and distance queries via 2-hop labels. \newblock {\em SIAM Journal on Computing}, 32(5):1338--1355, 2003.

\bibitem{dgpw-rdqmn-14} Daniel Delling, Andrew~V. Goldberg, Thomas Pajor, and Renato~F. Werneck. \newblock Robust distance queries on massive networks. \newblock In {\em Proceedings of the 22nd Annual European Symposium on Algorithms (ESA'14)}, volume 8737 of {\em Lecture Notes in Computer Science}, pages 321--333. Springer, September 2014.

\bibitem{dkp-pcbcp-12} Daniel Delling, Bastian Katz, and Thomas Pajor. \newblock Parallel computation of best connections in public transportation networks. \newblock {\em ACM Journal of Experimental Algorithmics}, 17(4):4.1--4.26, July 2012.

\bibitem{dpw-rbptr-14} Daniel Delling, Thomas Pajor, and Renato~F. Werneck. \newblock Round-based public transit routing. \newblock {\em Transportation Science}, 2014. \newblock accepted for publication.

\bibitem{dpsw-isftr-13} Julian Dibbelt, Thomas Pajor, Ben Strasser, and Dorothea Wagner. \newblock Intriguingly simple and fast transit routing. \newblock In {\em Proceedings of the 12th International Symposium on Experimental Algorithms (SEA'13)}, volume 7933 of {\em Lecture Notes in Computer Science}, pages 43--54. Springer, 2013.

\bibitem{d-ntpcg-59} Edsger~W. Dijkstra. \newblock A note on two problems in connexion with graphs. \newblock {\em Numerische Mathematik}, 1:269--271, 1959.

\bibitem{g-ctnrt-10} Robert Geisberger. \newblock Contraction of timetable networks with realistic transfers. \newblock In {\em Proceedings of the 9th International Symposium on Experimental Algorithms (SEA'10)}, volume 6049 of {\em Lecture Notes in Computer Science}, pages 71--82. Springer, May 2010.

\bibitem{jw-sfsro-13} Ruoming Jin and Guan Wang. \newblock Simple, fast, and scalable reachability oracle. \newblock {\em Proceedings of the VLDB Endowment}, 6(14):1978--1989, 2013.

\bibitem{ms-prcha-14} Florian Merz and Peter Sanders. \newblock Preach: A fast lightweight reachability index using pruning and contraction hierarchies. \newblock In {\em Proceedings of the 22nd Annual European Symposium on Algorithms (ESA'14)}, volume 8737 of {\em Lecture Notes in Computer Science}, pages 701--712. Springer, September 2014.

\bibitem{mw-op-06} Matthias {M{\"u}ller--Hannemann} and Karsten Weihe. \newblock On the cardinality of the {P}areto set in bicriteria shortest path problems. \newblock {\em Annals of Operations Research}, 147(1):269--286, 2006.

\bibitem{pswz-emtip-08} Evangelia Pyrga, Frank Schulz, Dorothea Wagner, and Christos Zaroliagis. \newblock Efficient models for timetable information in public transportation systems. \newblock {\em ACM Journal of Experimental Algorithmics}, 12(2.4):1--39, 2008.

\bibitem{sabw-ferra-13} Stephan Seufert, Avishek Anand, Srikanta Bedathur, and Gerhard Weikum. \newblock Ferrari: Flexible and efficient reachability range assignment for graph indexing. \newblock In {\em Proceedings of the 29th International Conference on Data Engineering}, pages 1009--1020. IEEE Computer Society, 2013.

\bibitem{sw-csa-13} Ben Strasser and Dorothea Wagner. \newblock Connection scan accelerated. \newblock In {\em Proceedings of the 16th Meeting on Algorithm Engineering and Experiments (ALENEX'14)}, pages 125--137. SIAM, 2014.

\bibitem{yaiy-fsrqg-13} Yosuke Yano, Takuya Akiba, Yoichi Iwata, and Yuichi Yoshida. \newblock Fast and scalable reachability queries on graphs by pruned labeling with landmarks and paths. \newblock In {\em Proceedings of the 22nd International Conference on Information and Knowledge Management}, pages 1601--1606. ACM Press, 2013.

\bibitem{ycz-grail-10} Hilmi Yildirim, Vineet Chaoji, and Mohammad~J. Zaki. \newblock Grail: Scalable reachability index for large graphs. \newblock {\em Proceedings of the VLDB Endowment}, 3(1):276--284, 2010.

\bibitem{zlwx-rqldg-14} Andy~Diwen Zhu, Wenqing Lin, Sibo Wang, and Xiaokui Xiao. \newblock Reachability queries on large dynamic graphs: A total order approach. \newblock In {\em Proceedings of the 2014 ACM SIGMOD International Conference on Management of Data (SIGMOD'14)}, pages 1323--1334. ACM Press, 2014.

\end{thebibliography}

\end{small}

\end{document}